\documentclass[12pt]{article}
\usepackage{amsmath}
\usepackage{amssymb}

\newcommand{\re}{\mathsf{Re}}

\newcommand{\opr}[1]{\mathsf{#1}}
\newcommand{\laplace}{\triangle}
\newcommand{\grad}{\nabla}
\newcommand{\pd}{\partial_{t_2}}
\newcommand{\pt}{\partial_{t_3}}
\renewcommand{\d}{\mathsf{d}}
\renewcommand{\[}{\begin{equation}}
\renewcommand{\]}[1]{\label{eq:#1}\end{equation}}

\newenvironment{proof}{Proof:}{$\blacksquare$}

\newtheorem{defin}{Definition}
\newtheorem{lemma}[defin]{Lemma}
\newtheorem{theorem}[defin]{Theorem}

\begin{document}

\begin{flushleft}

\textbf{\Large A Remark on Helical Waveguides} \\
[2em]
{\large Pavel Exner and Martin Fraas} \\ [.2em]
{\small \emph{Nuclear Physics Institute, Czech Academy of
Sciences,
25068 \v{R}e\v{z} near Prague, \\
Doppler Institute, Czech Technical University, B\v{r}ehov\'{a} 7,
11519 Prague, Czechia \\ e-mail: exner@ujf.cas.cz,
fraas@ujf.cas.cz}}
\\ [1em]
\end{flushleft}
{\small \textbf{Abstract.} Motivated by a proposal to create an
optical helix-shaped wave\-guides for cold atoms and molecules, we
discuss local perturbations which can create bound states in such
a setting. This is known about a local slowdown of the twist; we
show that a similar effect can result from a local tube protrusion
or a change of the helix radius in correlation with its pitch
angle.}

\vspace{3em}

\setcounter{equation}{0}
\section{Introduction}

Geometrically induced bound states in tubular regions appear in
various contexts -- see, e.g., \cite{DucEx, Chenaud} and
references therein. In particular, effects of waveguide twisting
were discussed recently in \cite{Krej, ExKov}. An independent
strong motivation to look into such a problem came from a study
\cite{Bha} which showed how helical waveguides for cold atoms and
molecules can be created by means of two counter-propagating
beams. Such waveguides do not have hard walls, of course, but it
is known that spectrum-generating geometric mechanisms are robust
-- see, e.g., \cite{leaky} -- so it is useful to study such
systems in a model which allows a mathematical treatment.

It was observed in \cite{ExKov} that slowing down the waveguide
twist acts as an effective attractive potential. Since the
mentioned optical waveguides allow for various modifications it is
useful to look for other perturbations which could have a similar
effect. We will discuss two of them. One is a \emph{local
protrusion} which is well known to create bound states in straight
tubes \cite{Simon}, however, the proof needs to modified to fit to
helical waveguides.

Another possible perturbation is a variation of the helix radius.
Here it is more suitable to employ the formalism introduced
originally in \cite{DucEx}. An exhaustive analysis would be
complicated, however, and we restrict here to the situation of a
thin waveguide when the cross section size is much smaller than
the helix radius, paying attention to particular cases, those of a
circular channel and of a flat ``ribbon-shaped'' one. In both
cases we observe that a local \emph{reduction} of the helix radius
can induce a weakly bound state if the ``pitch angle'' is large
enough, and in contrast, at small pitch angles localized states
may appear for a local radius \emph{enlargement}. At the same
time, the crossover between the two regimes depends on the cross
section.

\setcounter{equation}{0}
\section{Helical channel Hamiltonians}

We will employ the notation of \cite{ExKov} which we extend to the
case of varying cross-section and helix radius. Let $\omega$ be an
open connected set in $\mathbb{R}^2$ we fix $t^0=(t_2^0,\,t_3^0)
\in \omega$ and define
\[\omega(\alpha) = \left\{l_\alpha(t):= \left(\alpha(t_2 - t_2^0)
+ t_2^0,\,\alpha (t_3 - t_3^0) + t_3^0 \right):\: t \in \omega
\right\};
\]{omegaa}
changing $\alpha$ means a radial scaling of the cross section
$\omega$ w.r.t. the point $t^0$. Let further $\theta(s),
\alpha(s)$ be differentiable functions from $\mathbb{R}$ to
$\mathbb{R}$. We define the mapping $\mathcal{L}$ from $\mathbb{R}
\times \omega$ to $\mathbb{R}^3$ by $\mathcal{L} :=
\mathcal{\tilde{L}}(s, \,\tilde t)$, where tilde\footnote{It would
be appropriate, of course, to use $t$ indexed by the scaling
parameter and to mark its dependence on $s$. Such a notation,
however would be rather cumbersome and we believe that no
confusion will arise. The tilde will be later used for functions
on the scaled cross section $\omega(\alpha)$ as well.} marks the
scaling (\ref{eq:omegaa}) for $\alpha := \alpha(s)$, i.e. $\tilde
t = (\tilde t_2,\,\tilde t_3) =\big(\alpha(s)(t_2 - t_2^0) +
t_2^0,\,\alpha(s) (t_3 - t_3^0) + t_3^0 \big)$, and
\[
\mathcal{\tilde{L}}(s,\,t) :=\big(s,\,t_2 \cos \theta(s) + t_3
\sin \theta(s),\,t_3 \cos \theta(s) - t_2 \sin \theta(s) \big).
\] {maping}
We denote the $\mathcal{L}$-image of $\mathbb{R}\times\omega$ by
$\Omega$. It is tube in $\mathbb{R}^3$ which is purely helical if
the functions $\dot\theta(s), \alpha(s)$ are constant. We will be
concerned with a quantum particle concerned within this tube
assumimg that it has hard walls. Using appropriate units to get
rid of unnecessary constants, we can then identify the particle
Hamiltonian with the Dirichlet Laplacian $\opr{H}$ on
$L^2(\Omega)$, i.e. the self-adjoint operator associated with the
closed quadratic form
\[
\opr{Q}[\psi] := \int_\Omega|\grad \psi|^2\, \d t\, \d s, \quad
\psi \in \mathcal{H}^1_0(\Omega).
\] {q1}
The family of operators\footnote{The function $\theta$ will be
kept fixed so we do not write it explicitly in the operator
symbol.} corresponding to given $\theta(s), \alpha(s)$ will be
written as $\opr{H}(\alpha)$. In particular, $\opr{H}_0(\alpha)$
will denote the operator corresponding to $\theta(s)=\beta_0
s,\,\alpha(s)$, and $\opr{H_0}$ will refer to the purely helical
tube, $\theta(s) = \beta_0 s,\,\alpha(s) = \alpha_0$. A simple
substitution of variables shows that the quadratic form
$Q_0(\alpha)$ associated with $\opr{H}_0(\alpha)$ is unitarily
equivalent to
\[
\opr{Q}_0(\alpha)[\psi] = \int_{\mathbb{R}}
\int_{\omega(\alpha(s))}( |\grad_t \psi|^2 + |\partial_s \psi +
\beta_0 \psi'_\tau|^2) \d t\, \d s,
\] {QuadForm}
denoted for simplicity by the same symbol, where we have
introduced the notation $f'_\tau := (t_2 \pt - t_2 \pd)f$ in
accordance with \cite{ExKov}.

Furthermore, the quadratic form (\ref{eq:q1}) can transformed to
other unitarily equivalent expressions supported by a straight
tube (cylinder) is such a way that the geometric information is
contained in the coefficients. Following \cite{Krej} we arrive at
the quadratic form $\opr{q}' $ \cite{Krej} acting on
$L^2(\Omega_0,|G|^{1/2})$ as
\[
\opr{q}'[\psi]:=\int_{\mathbb{R} \times
\omega}\overline{(\partial_i \psi)} G^{ij} (\partial_j \psi)
|G|^{1/2} \,\d s \,\d t,
\] {tilq}
where we number the variables in such a way that
$(\partial_1,\,\partial_2,\,\partial_3) =
(\partial_s,\,\pd,\,\pt)$. By a straightforward computation we get
for the metric tensor the expression
\[
G^{ij} = \left( \begin{array}{ccc}
                    1 & - \frac{h_2}{\alpha(s)} & -\frac{h_3}{\alpha(s)
                    } \\
                    -\frac{h_2}{\alpha(s)} & \frac{1+h_2^2}{\alpha^2(s)} &
                    \frac{h_2 h_3}{\alpha^2(s)} \\
                    -\frac{h_3}{\alpha(s)} & \frac{h_2 h_3}{\alpha^2(s)
                    } & \frac{1 + h_3^2}{\alpha^2(s)}
                \end{array} \right),
\] {metric}
where
\begin{eqnarray}
& h_2 = (t_2 - t_2^0) \dot{\alpha}(s) + \tilde t_3 \dot\theta(s), \\
& h_3 = (t_3 - t_3^0) \dot \alpha(s) -\tilde t_2 \dot \theta(s).
\end{eqnarray}
We can also pass to a quadratic form on the Hilbert space
$L^2(\mathbb{R} \times \omega)$ without the additional weight
$|G^{1/2}|$; this is achieved by putting \cite{Krej}
\begin{eqnarray}
\lefteqn{\opr{q}[\psi]:=\opr{q}'[|G|^{-1/4}\psi]} \nonumber \\ &&
= (\partial_i \psi,\,G^{ij}\partial_j \psi) + (\psi,\,(\partial_i
F)G^{ij}(\partial_j F) \psi) - 2 \re (\partial_i \psi,\, G^{ij}(
\partial_j F) \psi), \label{q3}
\end{eqnarray}
where $F:=\log(|G|^{1/4})$. If we finally plug in the metric
tensor (\ref{eq:metric}) we get
\begin{multline}
\opr{q}[\psi] = \int \frac{1}{\alpha^2(s)} (|\alpha(s) \partial_s
\psi - h_2 \pd \psi - h_3 \pt \psi -\dot{\alpha}(s) \psi|^2 +
|\grad_t \psi|^2) \,\d s \,\d t. \label{eq:cq}
\end{multline}

After these preliminaries we shall formulate a theorem which will
be proved in Section~\ref{sec:bumps} below. We will consider
helical tubes with a constant ``pitch angle''$\!$, $\theta(s) =
\beta_0 s\:$ for some $\beta_0>0$. Furthermore, we assume that
outside a compact region $\Omega$ is purely helical; without loss
of generality we may suppose that $\alpha(s)=1$ there so the
unperturbed tube cross section is $\omega$.

\begin{theorem}
\label{def:t1} Suppose that $\alpha(s)-1$ is a nonzero continuous
function, which is nonnegative and compactly supported. Let
further $\omega,\, t^0$ be such that $\omega(\alpha) \subset
\omega(\alpha')$ holds for $\alpha\le\alpha'$, then the operator
$\opr{H}_0(\alpha)$ has at least one eigenvalue below the
threshold of the essential spectrum.
\end{theorem}
\textbf{Remark} We exclude here the case of a straight tube,
$\beta_0=0$. Bound states induced by a local protrusion exists in
such a situation also, however, one can use a more straightforward
way to prove the claim -- cf.~\cite{Simon}.

\setcounter{equation}{0}
\section{Spectrum of $\opr{H}_0$}

As in \cite{ExKov} our strategy is to regard $\opr{H}_0(\alpha)$
as a perturbation of $\opr{H}_0=\opr{H}_0(1)$. The spectrum of the
latter operator is purely absolutely continuous and covers the
half-line $[E(1),\,\infty)$, where $E(\alpha)$ is the lowest
eigenvalue of the operator
\[
\tilde{\opr{h}}(\alpha)  = - \laplace^{\omega(\alpha)}_D -
\beta_0^2 (t_2 \pt - t_3\pd)^2
\] {CrossOp}
acting in $L^2(\omega(\alpha))$ (pay attention to the fact that
this $\tilde{\opr{h}}(1)$ corresponds to $\opr{h}(0)$ of
\cite{ExKov}). Moreover, the ground state is non-degenerate and
the corresponding ground-state eigenfunction $\tilde f_\alpha(t)$
is strictly positive in $\omega(\alpha)$.

A simple substitution of variables given by $l_\alpha$
(\ref{eq:omegaa}) allows us to pass to the unitarily equivalent
operator $\opr{h}(\alpha)$ acting on $L^2(\omega)$, thus without a
tilde. It acts as follows,
\begin{multline}
(\psi,\,\opr h(\alpha)\psi) = \frac{1}{\alpha^2}(\psi,\, \opr h(1)
\psi) -\frac{\beta_0^2}{\alpha^2} \int_\omega \Big[ |t_2\pt\psi -
t_3\pd\psi|^2 \\ - |(\alpha(t_2 - t_2^0) + t_2^0)\pt\psi -
(\alpha(t_3 - t_3^0) + t_3^0)\pd\psi|^2 \Big] \d t.
\label{eq:OprAlpha}
\end{multline}

Next we ask about the dependence of $\opr{h}(\alpha)$ on $\alpha$.
By $f_\alpha(t)$ we denote the normalized ground-state eigenvector
of $\opr{h}(\alpha)$, i.e. $\tilde f_\alpha(l_\alpha( t)) =
f_\alpha(t)$.

\begin{lemma}
\label{def:l1}
 $f_\alpha(t),\,E(\alpha)$ are real-analytic functions of $\alpha$
 in $(0,\infty)$. In particular, there are $E^{(1)}<0$ and $f^{(1)}(t)$
 such that $E(\alpha) = E + (\alpha -1)E^{(1)}+o(\alpha -1)$
 and $f_\alpha(t) = f(t) + (\alpha -1)f^{(1)}(t) + o(\alpha -1)$.
\end{lemma}

\noindent
\begin{proof}
We rewrite and estimate the integral on the right-hand side of
(\ref{eq:OprAlpha}),
\begin{multline}
\frac{\beta_0^2}{\alpha^2}\int_\omega \Big[ -|t_2\pt\psi -
t_3\pd\psi|^2 \\ +|\alpha(t_2 \pt \psi - t_3 \pd \psi) +
(1-\alpha)(t_2^0 \pt \psi - t_3^0 \pd) \psi|^2 \Big] \,\d t
\\ \leq \frac{\beta_0^2}{\alpha^2}|1-2\alpha^2|\int_\omega|t_2\pt \psi -
t_3\pd \psi|^2 \,\d t \\ \phantom{AAA} +
4\frac{\beta_0^2}{\alpha^2}\, (1-\alpha)^2\int_\omega(|t_2^0 \pt
\psi|^2 + |t_3^0\pd \psi|^2)\,\d t \\ \le C_1 (\psi,\,\opr
h(1)\psi) \nonumber
\end{multline}
for some $C_1>0$, since the last integral can be estimated by
$\,(\mathrm{diam\,}\omega\, \|\nabla_t\psi\|)^2$. It follows that
the operators $\opr{h}(\alpha)$ form an analytic family type of
type (B) -- cf.~\cite[Sec.~VII.4]{Kato}, note that it is a
particular case of exercise 4.23. there. Hence the analyticity of
$f_\alpha,\,E(\alpha)$ follows from finite multiplicity of the
ground state; recall that in fact it is non-degenerate.

It remains to prove that $E^{(1)}<0$ which we will do using the
minimax principle. Suppose that $\alpha > 1$ and consider the test
function obtained as a shifted ground state, $\tilde\psi(t) := f(
t_2 + (\alpha-1) a_2,\, t_3 + (\alpha-1) a_3)$, where $f$ means
here the lowest eigenfunction of $\opr{h}(1)$ extended to
$\omega(\alpha)$ by zero. For small enough shifts $a_2,\,a_3$ we
have $\{t_2 + (\alpha-1) a_2,\,t_3 + (\alpha-1) a_3\} \subset
\omega(\alpha)$ for all $t=(t_1,\,t_2) \in \omega$, and
consequently
\begin{multline}
(\tilde \psi,\,\tilde{\opr{h}}(\alpha)\tilde \psi)
= ( f,\,\opr{h}(1)f)  \\ + 2 (\alpha-1)  \beta_0^2
\int_\omega (t_2\pt f - t_3 \pd f)(-a_2 \pt f + a_3 \pd f) \,\d t \\
+ (\alpha-1)^2 \int_\omega  |a_2 \pt f - a_3 \pd f|^2 \,\d t.
\end{multline}
If the term linear in $(\alpha -1)$ does not vanish identically we
are done. Suppose, on the contrary, that
 $$
 \int_\omega (t_2 \pt f - t_3 \pd f) \pt f \,\d t =
 \int_\omega (t_2 \pt f - t_3 \pd f) \pd f \,\d t =0
 $$
holds, then we employ another test function, for instance one
obtained by scaling, $\tilde\psi(t) := f
\left(\frac{t_2-t_2^0}{\alpha} + t_2^0,\, \frac{t_3 -
t_3^0}{\alpha} + t_3^0 \right)$. Using (\ref{eq:OprAlpha}) we get
\begin{multline}
  ( \tilde \psi,\,\tilde{\opr{h}}(\alpha)\tilde \psi)
  = E - \beta_0^2 \int_\omega
  |t_2 \pt f - t_3 \pd f|^2 \,\d t - (1-\frac{1}{\alpha^2})
  \int_\omega|\grad_t f|^2 \,\d t \\+ \beta_0^2 \int_\omega
  |t_2 \pt f - t_3 \pd f|^2 \,\d t +(\alpha-1)^2
  \frac{\beta_0^2}{\alpha^2} \int_\omega|t_2^0 \pt f - t_3^0 \pd
  f|^2 \,\d t \\
  =E - 2(\alpha -1)\int_\omega |\grad_t f|^2 \,\d t + o(\alpha
  -1),
  \nonumber
\end{multline}
and since $\int_\omega |\grad_t f|^2 \,\d t > 0$ we get the sought
result.
\end{proof}

\setcounter{equation}{0}
\section{Helical channel with a protrusion}
\label{sec:bumps}

Now we are in position to prove Theorem~\ref{def:t1}. By
assumption the function $\alpha(s)-1$ is compactly supported,
hence $\opr{H}_0(\alpha) - \opr{H}_0$ is relatively compact and
$\sigma_{ess}(\opr{H}_0(\alpha)) = \sigma_{ess}(\opr{H}_0)$. Since
we know the essential spectrum threshold, we can find eigenvalues
below it using a variational estimate.

Since $\alpha(s)-1$ is supposed to be nonzero and non-negative,
one can find an interval $(-s_0,\,s_0)$ within the support of this
function on which the inequality $1 + \varepsilon |s - s_0|
<\alpha(s)$ holds for $\varepsilon$ small enough. It follows from
the domain monotonicity of Dirichlet Laplacian that it is
sufficient to establish existence of a bound state for $\alpha(s)
:=1 + \varepsilon|s-s_0|\chi(-s_0,s_0)$. As usual in such cases we
start constructing a trial function from then threshold-resonance
of $\opr{H}_0$. Given $\delta>0,\,\varepsilon >0 $ we put
$\Psi_{\delta,\,\varepsilon}(s,\,t) = f_\varepsilon(t)
\phi_\delta(s)$, where $f_\varepsilon$ is the ground-state
eigenfunction of $\opr{h} (1 + \varepsilon|s-s_0|\chi(-s_0,s_0))$
and
$$
\phi_\delta(s) = \left\{ \begin{array}{lcr}
                        e^{\delta(s+s_0)} & \mbox{if} & s \leq
                        -s_0, \\
                            1& \mbox{if} & -s_0 \leq s \leq s_0,
                            \\
                            e^{-\delta(s-s_0)} & \mbox{if} & s >
                            s_0.
                    \end{array} \right.
$$
We plug this expression into (\ref{eq:cq}) and by a
straightforward computation we get
\begin{multline}
  \opr{q}[\Psi_{\delta,\,\varepsilon}]
  - E||\Psi_{\delta,\,\varepsilon}||^2 = \delta +
  \int_{-s_0}^{s_0} \big(E(1 + \varepsilon |s-s_0|) - E(1)
  \big)\, \d s \\
  - 2 \re \int_{-s_0}^{s_0} \int_\omega  \frac{\beta_0}{\alpha(s)}
  (\tilde t_3 \pd f_\varepsilon -
  \tilde t_2 \pt f_\varepsilon) \\ \times
  \left(\partial_s f_\varepsilon - \frac{\dot \alpha(s)}{\alpha(s)}
  \Big( (t_2 - t_2^0)\pd f_\varepsilon + (t_3 - t_3^0)\pt f_\varepsilon -
  f_\varepsilon\Big) \right) \d s \,\d t  \\
  + \int_{-s_0}^{s_0} \int_\omega
  \left|\partial_s f_\varepsilon - \frac{\dot \alpha(s)}{\alpha(s)} \Big( (t_2 -
  t_2^0)\pd f_\varepsilon + (t_3 - t_3^0)\pt f_\varepsilon -
  f_\varepsilon \Big) \right|^2\, \d s \,\d t.
\end{multline}
Now we inspect the behavior of the expression for small
$\varepsilon$ using Lemma \ref{def:l1}. Using $\dot \alpha(s) \sim
\varepsilon \mathrm{sgn}(s)$ and $\partial_s f_\varepsilon \sim
\varepsilon \mathrm{sgn}(s) f^{(1)}$ we obtain
\begin{multline}
\opr{q}[\Psi_{\delta,\,\varepsilon}] -
E||\Psi_{\delta,\,\varepsilon}||^2 = \delta +
  \varepsilon \int_{-s_0}^{s_0} E^{(1)} |s-s_0| \,\d s  \\ - 2
  \varepsilon \beta_0 \int_{-s_0}^{s_0}
  \int_{\omega} (t_3^0 \pd f - t_2^0 \pt f)\,
  \mathrm{sgn}(s) \\ \times \left(f^{(1)} - (t_2 - t_2^0)
  \pd f - (t_3 - t_3^0) \pt f -f \right) \,\d t\,
  \d s + o(\varepsilon).
\end{multline}
The integral term vanishes due to the parity of the
$\mathrm{sign}$ function; putting then $\delta = \varepsilon^2$ we
arrive at
$$
\opr{q}[\Psi_{\delta,\,\varepsilon}] -
E||\Psi_{\delta,\,\varepsilon}||^2 =
  \varepsilon E^{(1)} s_0^2 +
  o(\varepsilon).
$$
Since $E^{(1)}$ is negative, it is sufficient to choose
$\varepsilon$ small enough to conclude the proof.

\setcounter{equation}{0}
\section{Thin helical tubes}

Next we look what happens if it is the radius rather than the
cross section of the helical tube which is locally changed; for
simplicity we restrict ourselves to small perturbations of a
\emph{thin} tube. To this aim it is more convenient to use the
approach due to \cite{GoldJ} and \cite{DucEx} where the cross
section is taken perpendicular to the tube axis. Let us stress
that while helical tubes do not fall into the class of the
asymptotically straight ones for which the existence of
geometrically induced discrete spectrum was established in
\cite{DucEx}, the perturbation theory w.r.t. the tube radius
developed there and in \cite{ClaBra} remains nevertheless valid
and we can use it here.

The generating curve of our thin channel will thus helix a varying
radius,
 $$
   \tilde\Gamma(t) = (t,\,R(t) \cos \theta(t),\,R(t) \sin
   \theta(t))\,,\quad t \in \mathbb{R}\,,
 $$
whose image in $\mathbb{R}^3$ will be denoted $[\Gamma]$. With
$\tilde\Gamma$ we conventionally associate its Frenet triad frame
$(\mathbf t,\, \mathbf n,\,\mathbf b)$ consisting of its tangent,
normal, and binormal vectors. Furthermore, $\kappa,\,\tau$ will
denote the curvature and torsion of $\tilde\Gamma$, respectively.
We suppose that
 $$ \theta(t) = \beta_0 t$$
and the radius is slightly changing according to
\begin{equation}  R(t) = R_0 + \varepsilon \delta(t)\,,
\label{eq:per}
\end{equation}
where $\delta(t)$ is a fixed (nonzero) $C^2$ smooth function of
compact support and $\varepsilon$ is a small parameter by which we
mean that $\varepsilon\|\delta\|_\infty \ll R_0$.

For further reference, let us first inspect the unperturbed helix,
$\varepsilon =0\,$; the correspondent quantities will be indicated
by the zero subscript. It is straightforward to check that for
$\Gamma_0(t) = (t,\,R_0 \cos \beta_0 t,\,\sin \beta_0 t)$ the
Frenet triad is
\begin{eqnarray}
  &&\mathbf t_0(t) =\left(\frac{1}{\sqrt{1+R_0^2\beta_0^2}},\,
  -\frac{R_0 \beta_0 \sin \beta_0 t}{\sqrt{1+R_0^2\beta_0^2}},\,
  \frac{R_0 \beta_0 \cos \beta_0 t}{\sqrt{1+R_0^2\beta_0^2}}\right),  \nonumber \\
  &&\mathbf n_0(t) = (0,\,-\cos \beta_0 t,\,-\sin \beta_0 t),  \nonumber \\
  &&\mathbf b_0(t) = \left(\frac{R_0 \beta_0}{\sqrt{1+R_0^2\beta_0^2}},\,
  \frac{\sin \beta_0 t}{\sqrt{1+R_0^2\beta_0^2}},\,
  -\frac{\cos \beta_0 t}{\sqrt{1+R_0^2\beta_0^2}} \right).  \nonumber
\end{eqnarray}
The normal vector $\mathbf n_0(t)$ is perpendicular to the helix
axis while the tangent $\mathbf t_0(t)$ and binormal $\mathbf
b_0(t)$ contains with it nontrivial angles independent of $t$; the
curvature and torsion are also constant and equal to
 $$
 \kappa_0 = \frac{R_0 \beta_0^2}{1+R_0^2 \beta_0^2}\,, \quad
 \tau_0 = \frac{\beta_0}{1 + R_0^2 \beta_0^2}\,.
 $$

To use the above mentioned results \cite{DucEx} and \cite{ClaBra}
we have replace $t$ in the parametrization of $[\Gamma]$ by the
arc length of the curve, $\Gamma(s):=\tilde\Gamma(t(s))$, where
$t(s)$ is determined by the implicit equation
\begin{equation}
  s=\int_0^{t(s)} |\dot{\tilde\Gamma}(\tau)|\, \d \tau\,. \label{eq:length}
\end{equation}
The parametrization change makes, of course, little difference for
the unperturbed helix where the two are mutually proportional,
$t(s)= s(1 + R_0^2 \beta_0^2)^{-1/2}$.

We will consider two models of thin helix quantum waveguides
corresponding to different cross-sections. The latter will a
family $\omega(s)$ of bounded connected neighborhoods of zero,
typically obtained by rotations of a fixed $\omega$ smooth w.r.t.
$s\,$; the tube in question $\Omega \subset \mathbb{R}^3$ is then
defined as the image of the map  $s\mapsto \Gamma(s) + x_2 \mathbf
n + x_3 \mathbf b\,,\: (x_2,\,x_3)\in \omega(s)\,$, as $s$ runs
through $\mathbb{R}$. The Hamiltonian is again the Dirichlet
Laplacian on $\Omega$ denoted by $\opr{H}$ and the parametric
description of $\Omega$ makes it possible to replace it by a
unitary equivalent operator on the ``straightened'' waveguide. The
two models we will be interested in are the following:
\begin{enumerate}
  \item[(i)] The cross-section is two-dimensional, circular with
  $\Gamma$ at its centre, and perpendicular to the helix. In that
  case we can make $\omega$ fixed in the so-called Tang coordinate
  system which rotates around $\mathbf t$ w.r.t. the Frenet triad
  with the angular velocity $\tau$. In that case one achieves a
  full decoupling of the longitudinal and transverse coordinates
  in the ``straightening'' transformation, see \cite{DucEx} for
  details. The perturbation theory with respect to the circle radius
  developed there shows, in particular, that the bottom of the
  spectrum for a thin tube is determined -- after subtracting the
  continuum threshold energy -- by the one-dimensional
  Hamiltonian $T := -\frac{\d^2}{\d^2s} +
  V_\mathrm{eff}^\mathrm{circ}(s)$ with the effective potential
  \begin{equation}
    V_\mathrm{eff}^\mathrm{circ}(s) := -\frac{1}{4} \kappa(s)^2
    \label{eq:VeffC}.
  \end{equation}
  \item[(ii)] The optical waveguides which we use as a motivation
  \cite{Bha} are, however, far of a circular shape having a very
  elongated cross section the sizes of which in two principal
  directions may differ by as much as two orders of magnitude. In
  such a case it is appropriate to use an idealized description due
  to \cite{ClaBra} in which the cross section is a one-dimensional
  segment and $\Omega$ has thus form of a winding ribbon; in accordance
  with \cite{Bha} we suppose that the segment $\omega(s)$ is
  perpendicular to the helix axis. To achieve that, the function
  $\alpha$ describing the rotation of $\omega$ must be such that
  $(\mathbf n \cos\alpha - \mathbf b \sin \alpha)(s)$ is
  perpendicular the axis direction for any $s$, i.e. that the first
  component of this vector vanishes. By the analysis of \cite{ClaBra}
  the weak-coupling problem is again described in the leading order
  by a one one-dimensional operator in which the effective potential
  (\ref{eq:VeffC}) is replaced by
  \begin{equation}
    V_\mathrm{eff}^\mathrm{ribbon}(s) := -\frac{1}{4} \kappa^2(s)
    \cos(\alpha(s))^2 + \frac{1}{2}\left(\tau(s) - \dot\alpha(s)\right)^2.
    \label{eq:VeffS}
  \end{equation}
\end{enumerate}

Conditions for the existence of weakly bound states can be thus
deduced from well-known properties of effective operator
$-\frac{\d^2}{\d^2s} + V_\mathrm{eff}$ which has isolated
eigenvalues if\footnote{We leave out the critical case, $\int
(V_\mathrm{eff}(s)-E_0)\, \d s = 0$, since using the effective
operator we deal the leading order only.} the potential is
attractive in the mean, $\int_\mathbb{R}\,
(V_\mathrm{eff}(s)-E_0)\, \d s < 0$, where $E_0 := \lim_{|s| \to
\infty} V(s))$; in our case the limit obviously exists since the
curvature and torsion are constant outside a compact set.

Let us look now what the above condition gives for the described
geometries. Since the radius perturbation is weak by assumption,
for the arc-length of perturbed helix we get from
(\ref{eq:length}) the relations
 $$
  s = \sqrt{1 + R_0^2 \beta_0^2}\, t(s) + \mathcal{O}(\varepsilon)\,,
  \quad
  t(s) = \frac{s}{\sqrt{1 + R_0^2 \beta_0^2}} +
  \mathcal{O}(\varepsilon) =: t_0(s) + \mathcal{O}(\varepsilon)\,;
  \label{eq:ts}
 $$
since $\delta\in C^2$ by assumption, also the first two
derivatives of $t(s)$ and $t_0(s)$ coincide up to
$\mathcal{O}(\varepsilon)$. Then we can compute the geometric
quantities which enter the above expressions for effective
potentials; after a straightforward if tedious computation we get
\begin{eqnarray*}
  && \kappa(s) = \kappa_0 + \frac{(\beta_0^2 - R_0^2
  \beta_0^4)\delta(t_0(s)) - (R_0^2 \beta_0^2 + 1)
  \ddot\delta(t_0(s))}{(1 + R_0^2 \beta_0^2)^2}\,\varepsilon
  + \mathcal{O}(\varepsilon^2)\,, \\
  &&\tau(s) = \tau_0 - 2 \frac{ R_0^2
  \beta_0^4 \delta(t_0(s)) + (R_0^2 \beta_0^2 + 1)
  \ddot\delta(t_0(s))}{R_0 \beta_0 (1 + R_0^2 \beta_0^2)^2}\,\varepsilon
  + \mathcal{O}(\varepsilon^2), \\
  && \tan \alpha(s) = -\frac{\dot\delta(t_0(s))}{R_0 \beta_0 \sqrt{1 +
  R_0^2 \beta_0^2}}\,\varepsilon + \mathcal{O}(\varepsilon^2)\,.
\end{eqnarray*}
Comparing this results with the effective potential for the ribbon
(\ref{eq:VeffS}) we see that the terms linear in $\ddot
\delta(t_0(s))$ do not contribute to the integral in the condition
$\int (V_\mathrm{eff}(s)-E_0)\, \d s = 0$ if $\delta$ is smooth as
assumed, and thus the ribbon twisting described by the function
$\alpha$ plays no role in the leading order. Computing the
effective potentials explicitly with the help of the above
formulae we arrive at the following conclusions:
\begin{enumerate}
\item[(i)] for the circular tube we have
\begin{multline}
 V_\mathrm{eff}^\mathrm{circ}(s) = - \frac{R_0^2 \beta_0^4}{4(1+R_0^2
 \beta_0^2)^2} \\ + \frac{R_0 \beta_0^4(R_0^2 \beta_0^2 -1) \delta(t(s))
 + R_0 \beta_0^2(R_0^2 \beta_0^2 + 1) \ddot\delta(t(s))}{2(1+R_0^2
 \beta_0^2)^3} \,\varepsilon + \mathcal{O}(\varepsilon^2)\,.
\end{multline}
Hence we can distinguish two cases: for a ``steep'' helix, $R_0
\beta_0>1$, a weakly bound state occurs if $\int_\mathbb{R}
\delta(t)\, \d t < 0$ , i.e. in the situation where the the helix
radius is \emph{locally reduced}. On the other hand, for a small
pitch angle, $R_0 \beta_0 < 1$ the bound state occurs if
$\int_\mathbb{R} \delta(t)\, \d t > 0$, i.e. if the radius is
locally \emph{enhanced}.

\item[(ii)] in the ribbon case we have
\begin{multline}
V_\mathrm{eff}^\mathrm{ribbon}(s)= -\frac{R_0^2 \beta_0^4 - 2
\beta_0^2}{4(1+R_0^2 \beta_0^2)^2} \\ + \frac{R_0^2
\beta_0^4(R_0^2 \beta_0^2 -5) \delta(t(s)) + (1 + R_0^2
\beta_0^2)(R_0^2 \beta_0^2 - 2) \ddot\delta(t(s))}{2 R_0(1 + R_0^2
\beta_0^2)^3} \,\varepsilon \\+ \mathcal{O}(\varepsilon^2)\,,
\end{multline}
and again we have two cases differing from the previous situation
just by the critical value of the pitch angle. For $R_0 \beta_0
> \sqrt5$ the bound state occurs under the local ``squeezing'', while for
$R_0 \beta_0 < \sqrt5 $ we have to ``inflate'' the helix locally
to achieve binding.
\end{enumerate}

\subsection*{Acknowledgments}

We are indebted to Dr. Bhattacharya for informing us about the
results in \cite{Bha} prior to publication. The research was
supported in part by the Czech Academy of Sciences and Ministry of
Education, Youth and Sports within the projects A100480501 and
LC06002.


\end{document}